\documentclass[twocolumn,english,showpacs]{revtex4}
\usepackage{babel,amsmath,amssymb,dcolumn}
\usepackage[dvips]{graphics}
\usepackage{latexsym}

\usepackage{psfrag,fancyhdr,epsfig}
\def\imo{i}
\begin{document}

\title{Massive scalar field quasi-normal modes of higher dimensional black holes}

\author{Alexander Zhidenko}\email{zhidenko@fma.if.usp.br}

\affiliation{Instituto de F\'{\i}sica, Universidade de S\~{a}o Paulo \\
C.P. 66318, 05315-970, S\~{a}o Paulo-SP, Brazil}

\pacs{04.30.Nk,04.50.+h}

%
% Pacs used:
%  04.30.Nk Wave propagation and interactions
%  04.50.+h Gravity in more than four dimensions, Kaluza-Klein theory,
%           unified field theories; alternative theories of gravity

\begin{abstract}
We study quasi-normal spectrum of the massive scalar field in the $D$-dimensional black hole background.
We found the qualitatively different dependence of the fundamental modes on the field mass for $D\geq6$.
The behaviour of higher modes is qualitatively the same for all $D$.
Thus for some particular values of $\mu M^{1/(D-3)}$, where $\mu$ is the field mass and $M$ is the black hole mass,
the spectrum has two dominating oscillations with a very long lifetime.
We show that the asymptotically high overtones do not depend on the field mass.
In addition, we present the generalisation of the Nollert improvement of the continued fraction technique
for the numerical calculation of quasi-normal frequencies of $D$-dimensional black holes.
\end{abstract}

\maketitle

%%%%%%%%%%%%%%%%%%%%%%%%%%%%%%%%%%%%%%%%%%%%%%%%%%%%%%%%%%%%%%%%%%%%%%%%%%%%%%%

\section{Introduction}

One of the attractive solutions of the hierarchy problem is the consideration of our 4-dimensional universe
as a sub-manifold that is embedded in a higher dimensional space-time.
The models of type ADD (Arkani-Hamed, Dimopoulos and Dvali) \cite{ADD}
allow compactification radius of extra dimensions to be of macroscopic size \cite{compactify}.
Black holes whose horizon radius is much smaller than this characteristic length can be well described by solution for higher-dimensional static black hole \cite{Tangherlini}.
Such black holes could be produced at the next-generation particle colliders, probably at energies of very low order $\sim1TeV$ \cite{creation}.
Thus study of their properties is strongly motivated.

The very important black hole characteristic is its quasi-normal (QN) spectrum: the set of complex frequencies (see \cite{QNMsRev} for review).
The real parts of the QN modes are the frequency of oscillations, while the imaginary parts are their damping rates.
These oscillations dominate on the intermediately late times of a perturbation evolution near a black hole.
They do not depend on the initial perturbation, being defined by the black hole parameters only.

While the QN spectrum of different objects in four-dimensional \cite{quasi-4D} and higher dimensional \cite{quasi-D} cases
was extensively studied for massless fields of different spin and for the perturbation of the metric itself,
our knowledge about the massive field QN spectrum still has gaps.
Massive scalar field QN modes were studied only for four-dimensional black holes within WKB approximation \cite{massivescalar},
and later with continued fraction method of Leaver \cite{quasiresonance,qrc,mkerr}.
Massive vector QNMs were investigated for Schwarschild-(anti) de Sitter black holes \cite{massivevector}.
Massive Dirac QNMs were studied in \cite{massiveDirac}.

It was found that the behaviour of massive fields is quite different from that of the massless ones:
QN spectrum of massive fields has infinitely long-living modes, called quasi-resonances \cite{quasiresonance,massivevector}.
The late time behaviour of the the massive fields in black hole background is the same as in pure Minkowski space-time.
Asymptotically late time behaviour of them does not depend on their spin, mass and multipole number \cite{massivelatetime}.
Also massive fields demonstrate the so-called superradiant instability \cite{superradiant}, which is absent for massless fields.

This paper is dedicated to the study of the QN spectrum of the massive scalar field in a $D$-dimensional black hole background.
Note, that the scalar field is widely considered as a good model for physical fields when the effect of their spin is negligibly small.
It appears also in cosmological models, and, as one of the fundamental states in string theories.

The scalar field with the mass term can be also interpreted as a self-interacting scalar field within regime of small perturbations \cite{Hod}.
Also, the scalar field gains large effective mass when considered in models with extra dimensions of RS-type \cite{RS}.
We study the QN spectrum in this regime as well.

This paper is organized as follows:
in Sec.~\ref{sec.basic} we present basic formulae and the method of numerical calculation of the quasi-normal modes,
Sec.~\ref{sec.numerical} is devoted to our numerical results and their comparison with those obtained in other works.
Finally, in Sec.~\ref{sec.conclusions} we summarize the obtained results.

\section{Basic equations and numerical analysis}\label{sec.basic}

The metric of the Schwarzshild black hole in $D$-dimensions has the
form \cite{MyersPerry}:
\begin{equation}\label{metric}
ds^2= f(r) dt^2 -f^{-1}(r) dr^2 - r^2 d \Omega^{2}_{D-2},
\end{equation}
where
$$f(r)=1-\left(\frac{r_{0}}{r}\right)^{D-3}= 1-\frac{16 \pi G M}{(D-2) \Omega_{D-2} r^{D-3}}.$$
Here we used the quantities
$$\Omega_{D-2}=\frac{(2 \pi)^{(D-1)/2}}{\Gamma((D-1)/2)}, ~~ \Gamma(1/2) =
\sqrt{\pi},~~ \Gamma(z+1)=z \Gamma(z).$$

%Using the following factorial ansatz:
%$$\Psi=e^(-\imo\omega t)r^{1-D/2}R(r)S_{D-2},$$

After separation of angular and time variables, the radial part of the massive scalar field equation can be reduced to the form:
\begin{equation}\label{radial}
\left(f(r)\frac{d}{dr}f(r)\frac{d}{dr}+\omega^2-V(r)\right)R(r)=0,
\end{equation}
where
\begin{eqnarray}\label{potential}
\nonumber V(r) &=& f(r)\left(\mu^2+\frac{l(l+D-3)}{r^2} + \right. \\
\nonumber &&\left. + \frac{(D-2) (D-4)}{4 r^2} f(r) + \frac{D-2}{2 r} f'(r)\right),
\end{eqnarray}
$l$ is the overtone number, $\mu$ is the field mass.

By definition, the quasi-normal modes are eigenvalues of $\omega$ with the boundary conditions
which correspond to the outgoing wave at spatial infinity and the ingoing wave at the black hole horizon.
In 4-dimensional case \cite{quasiresonance}, these boundary conditions look like
\begin{eqnarray}
&R &\simeq C_+ \exp(\imo\chi r)r^{\imo r_0(\chi+\mu^2/2\chi)}\,, \quad {\rm as} \quad r \rightarrow \infty\,,
\label{inftyBC4D} \\[3mm]
&R & \simeq C_- (r-r_0)^{-\imo\omega}, \quad {\rm as} \quad r \rightarrow r_h\,,
\label{horizonBC4D}
\end{eqnarray}
where $\chi=\sqrt{\omega^2-\mu^2}$. The sign of $\chi$ should be chosen in order to remain in the same complex quadrant as $\omega$.

It turns out that the behaviour of the wave function in a higher dimensional background differs qualitatively from that in the 4-dimensional one.
For $D>4$ the leading term of $V(r)-\mu^2$ at infinity is of order $r^{-2}$, that is why the appropriate boundary conditions for $D>4$ have the form:
\begin{eqnarray}
&R &\simeq C_+ \exp(\imo\chi r)\,, \quad {\rm as} \quad r \rightarrow \infty\,,
\label{inftyBC} \\[3mm]
&R & \simeq C_- (r-r_0)^{-\imo\omega/(D-3)}, \quad {\rm as} \quad r \rightarrow r_h\,.
\label{horizonBC}
\end{eqnarray}
One should note, that these boundary conditions do not change the possibility of quasi-resonance existing (see \cite{qrc}).

To solve the equation (\ref{radial}), we use the continued fraction method \cite{Leaver:1985ax}.
Equation (\ref{radial}) has an irregular singularity at spatial infinity $r=\infty$ and $D-2$ regular singular points: $r=0$ and the roots of the equation $(r/r_0)^{D-3}=1$, one of them is the horizon at $r=r_0$.
Thus the appropriate Frobenius series have the following form:
\begin{equation}\label{singularity}
R(r)=\exp(i\chi r)\left(1-\frac{r_0}{r}\right)^{-\imo\omega/(D-3)} u\left(1-\frac{r_0}{r}\right)\,.
\end{equation}
where
\begin{equation}\label{Frobenius}
u(z)=\sum_{i=0}^\infty b_iz^i\,\qquad \left(z=1-\frac{r_0}{r}\right),
\end{equation}
is regular in the circle $|z|<1$ for $D\leq9$.
For higher $D$, some of the singularities appear in the unit circle and one has to continue the Frobenius series through some midpoints.
This technique was recently developed in \cite{Rostworowski:2006bp}, but in the present work we were limited by $D\leq9$.

Substituting (\ref{singularity}) into (\ref{radial}), one can
obtain a $(2D-5)$-term recurrence relation for the coefficients $b_i$
\begin{equation}\label{rrelation}
\sum_{j=0}^{min(2D-6,i)} c_{j,i}^{(2D-5)}(\omega)\,b_{i-j}=0,\quad
{\rm for}\,\,i>0\,.
\end{equation}
Each of the functions $c_{j,i}^{(2D-5)}(\omega)$ has an analytical
form in terms of $s$, $i$ and $\omega$, for a specific
dimensionality $D$ and each integer $0\leq j\leq 2D-6$.

We now decrease the number of terms in the recurrence relation
\begin{equation}\label{srcRE}
\sum_{j=0}^{min(k,i)}c_{j,i}^{(k+1)}(\omega)\,b_{i-j}=0
\end{equation}
by one, i.\ e.\ we find the $c_{j,i}^{(k)}(\omega)$, which satisfy
the equation
\begin{equation}\label{finRE}
\sum_{j=0}^{min(k-1,i)}c_{j,i}^{(k)}(\omega)\,b_{i-j}=0\,.
\end{equation}
For $i\geq k$, we can rewrite the above expression as
\begin{equation}\label{subsRE}
\frac{c_{k,i}^{(k+1)}(\omega)}{c_{k-1,i-1}^{(k)}(\omega)}
\sum_{j=1}^{k}c_{j-1,i-1}^{(k)}(\omega)\,b_{i-j}=0.
\end{equation}
Subtracting (\ref{subsRE}) from (\ref{srcRE}) we find the relation
(\ref{finRE}) explicitly. Thus we obtain:
\begin{eqnarray}
&&c_{j,i}^{(k)}(\omega) = c_{j,i}^{(k+1)}(\omega),\qquad
{\rm for}\,\,j=0,\,\,\mbox{or}\,\,i<k,\nonumber \\[2mm]
&&c_{j,i}^{(k)}(\omega) = c_{j,i}^{(k+1)}(\omega)-\frac{c_{k,i}^{(k+1)}(\omega)\,
c_{j-1,i-1}^{(k)}(\omega)}{c_{k-1,i-1}^{(k)}(\omega}\,.
\nonumber%\label{GaussElimination}
\end{eqnarray}
This procedure is called \emph{Gaussian eliminations}, and allows us
to determine the coefficients in the three-term recurrence relation
\begin{eqnarray}
&&c_{0,i}^{(3)}\,b_i+c_{1,i}^{(3)}\,b_{i-1}+c_{2,i}^{(3)}\,b_{i-2}=0, \quad
{\rm for}\,\,i>1\nonumber\\[1mm]
&&c_{0,1}^{(3)}\,b_1+c_{1,1}^{(3)}\,b_0=0,
\end{eqnarray}
numerically for a given $\omega$ up to any finite $i$. The
complexity of the procedure is \emph{linear} with respect to $i$ and $k$.

The requirement that the Frobenius series be convergent at spatial
infinity implies that
\begin{equation}
0=c_{1,1}^{(3)}-\frac{c_{0,1}^{(3)}c_{2,2}^{(3)}}{c_{1,2}^{(3)}-}
\,\frac{c_{0,2}^{(3)}c_{2,3}^{(3)}}{c_{1,3}^{(3)}-}\ldots\,,
\end{equation}
which can be inverted $n$ times to give
\begin{eqnarray}
&~&
c_{1,n+1}^{(3)}-\frac{c_{2,n}^{(3)}c_{0,n-1}^{(3)}}{c_{1,n-1}^{(3)}-}
\,\frac{c_{2,n-1}^{(3)}c_{0,n-2}^{(3)}}{c_{1,n-2}^{(3)}-}\ldots\,
\frac{c_{2,2}^{(3)}c_{0,1}^{(3)}}{c_{1,1}^{(3)}}=\nonumber\\
\label{CFeq}&~&=\frac{c_{0,n+1}^{(3)}c_{2,n+2}^{(3)}}{c_{1,n+2}^{(3)}-}
\frac{c_{0,n+2}^{(3)}c_{2,n+3}^{(3)}}{c_{1,n+3}^{(3)}-}\ldots\,.\label{invcf}
\end{eqnarray}
The equation (\ref{invcf}) with \emph{infinite continued fraction} on the
right-hand side can be solved numerically by minimising the
absolute value of the difference between the left- and right-hand
sides. The equation has an infinite number of roots (corresponding
to the QN spectrum), but for each $n$, the most stable root is
different. In general, we have to use the $n$ times inverted
equation to find the $n$-th QN mode. The requirement that the
continued fraction be itself convergent allows us to limit its
depth by some large value, always ensuring that an increase in
this value does not change the final results within the desired
precision.

It turns out, that the convergence of the infinite continued fraction becomes worse
if the imaginary part of $\omega$ increases with respect to the real part.
This problem was circumvented  by H.-P. Nollert \cite{Nollert}.
To improve the convergence of the infinite continued fraction one can consider initial approximation for
\begin{equation}\label{NollertR}
-\frac{b_n}{b_{n-1}}=R_n=\frac{c_{2,n+1}^{(3)}}{c_{1,n+1}^{(3)}-}
\frac{c_{0,n+1}^{(3)}c_{2,n+2}^{(3)}}{c_{1,n+2}^{(3)}-}\ldots\,,
\end{equation}
that for large $n$ can be expanded as
\begin{equation}\label{NollertExp}
R_n(\omega)=C_0(\omega)+\frac{C_1(\omega)}{\sqrt{n}}+\frac{C_2(\omega)}{n}+\ldots\,.
\end{equation}
Since for the four-dimensional case the coefficients $c_{0,n}^{(3)}$, $c_{1,n}^{(3)}$ and $c_{2,n}^{(3)}$ are known analytically,
one can find $C_i(\omega)$ from the equation
\begin{equation}\label{NollertEq}
R_n=\frac{c_{2,n+1}^{(3)}}{c_{1,n+1}^{(3)}-c_{0,n+1}^{(3)}R_{n+1}}.
\end{equation}

The found expansion (\ref{NollertExp}) could be used as an initial approximation for the "remaining" infinite continued fraction.
In order to ensure the convergence of (\ref{NollertExp}) for a given value of $\omega$,
one has to start from the found approximation deeply enough inside the continued fraction (\ref{CFeq}).
It turns out, that the required depth is less than it would be if we start from some arbitrary value as the initial approximation.

As mentioned in \cite{qrc}, quasi-resonances correspond to purely imaginary $\chi$. Thus
the convergence of the continued fraction decreases dramatically,
and one needs to use the described method for calculation of those frequencies.

Even though $c_{0,n}^{(3)}$, $c_{1,n}^{(3)}$ and $c_{2,n}^{(3)}$ are not known analytically in the case under consideration,
we are still able to find the expansion (\ref{NollertExp}) directly from (\ref{rrelation}). After dividing the equation (\ref{rrelation}) by
$b_{i-2D+6}$ and using the definition of $R_n=-b_n/b_{n-1}$, one can find
\begin{equation}\label{NollertHD}
\sum_{j=0}^{2D-6}(-1)^j\,c_{j,i}^{(2D-5)}(\omega)\,\prod_{k=0}^{2D-7-j}R_{i-k}=0,\quad
{\rm for}\,\,i\gg1\,.
\end{equation}

The important is that the expression (\ref{NollertExp}) should be defined without any indeterminations.
Since the infinite continued fraction is convergent, it is clear that $R_n$ is itself well-defined.
Our task is just to recover it from the equation (\ref{NollertHD}).
It turns out that only determination of $C_0(\omega)$ and $C_1(\omega)$ could cause ambiguities:
the first appears due to a lot of roots of the equation (\ref{NollertHD})
and the second is connected with two possible signs of $\sqrt{n}$ in (\ref{NollertExp}).
Fortunately, both ambiguities can be eliminated by using our knowledge about the properties of the series (\ref{Frobenius}).
Let us consider them more precisely.

For large $n$ $c_{j,n}^{(2D-5)}(\omega)\propto n^2$, thus $C_0(\omega)$ satisfies
\begin{equation}\label{C0eq}
\lim_{n\rightarrow\infty}\frac{1}{n^2}\sum_{j=0}^{2D-6}(-1)^j\,c_{j,n}^{(2D-5)}(\omega)\,C_0^{2D-6-j}(\omega)=0\,.
\end{equation}
The equation (\ref{C0eq}) has in general $2D-6$ roots (in fact there are repeated roots).
One of the roots is \textit{always} $C_0=-1$ (it is also repeated), implying the unit radius of convergence of the series (\ref{Frobenius}).
Other roots appear due to the existing of additional singular points of the equation (\ref{radial}).
Thus we choose $C_0=-1$.

After fixing $C_0=-1$ one can find that $$C_1=\pm\sqrt{-2\imo r_0\chi}.$$
The sign of $C_1$ can be recovered using the convergence of the series (\ref{Frobenius}) at spatial infinity ($z=1$).
Therefore:
$$\lim_{n\rightarrow\infty}b_n=0,$$
i.\ e.\ $\nexists N:~\forall n>N~|b_n|>|b_{n-1}|$.
Since for large $n$
$$\frac{b_n}{b_{n-1}}\sim -R_n\sim -C_0-\frac{C_1}{\sqrt{n}}=1-\frac{C_1}{\sqrt{n}}\,,$$
we find out that the real part of $C_1$ cannot be negative.

After the sign of $C_1$ is fixed, the other coefficients in (\ref{NollertExp}) can be found without encountering indeterminations \cite{Nollert}.

Thus we have now possibility to use (\ref{NollertExp}) as an initial approximation for $R_n$ when $n\gg r_0\chi$.
This allows us to improve the convergence of the infinite continued fraction near quasi-resonances.

%\begin{widetext}

\begin{figure*}
\begin{center}
\includegraphics[width=\textwidth, clip]{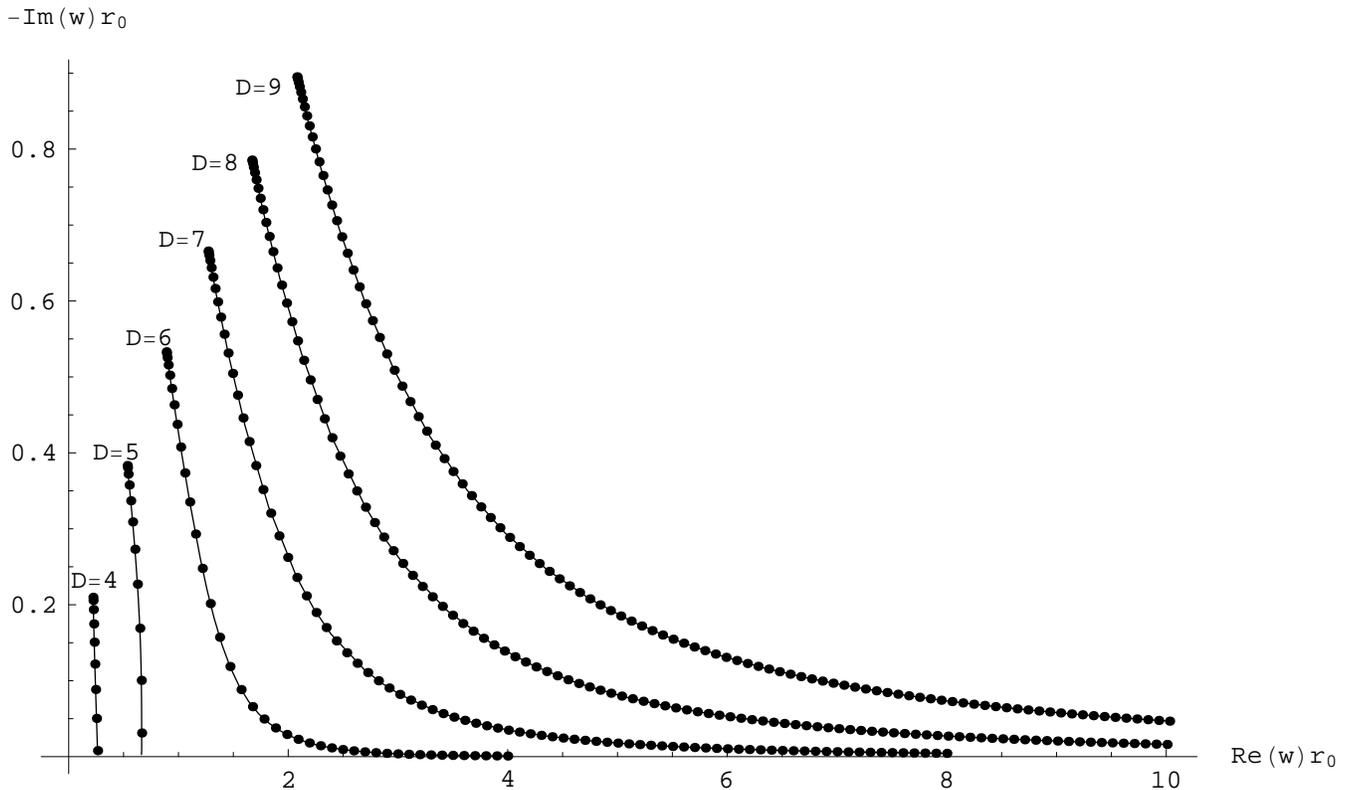}
\caption{Fundamental quasi-normal frequencies ($l=0$) for different $D$ and $\mu$.
The frequency for $\mu=0$ has the largest imaginary part. The points were plotted with the step of $\Delta\mu=0.1/r_0$.
Solid lines mark the same number of $D$.}
\label{fig.l=0}
\end{center}
\end{figure*}

\begin{figure*}
\begin{center}
\includegraphics[width=\textwidth, clip]{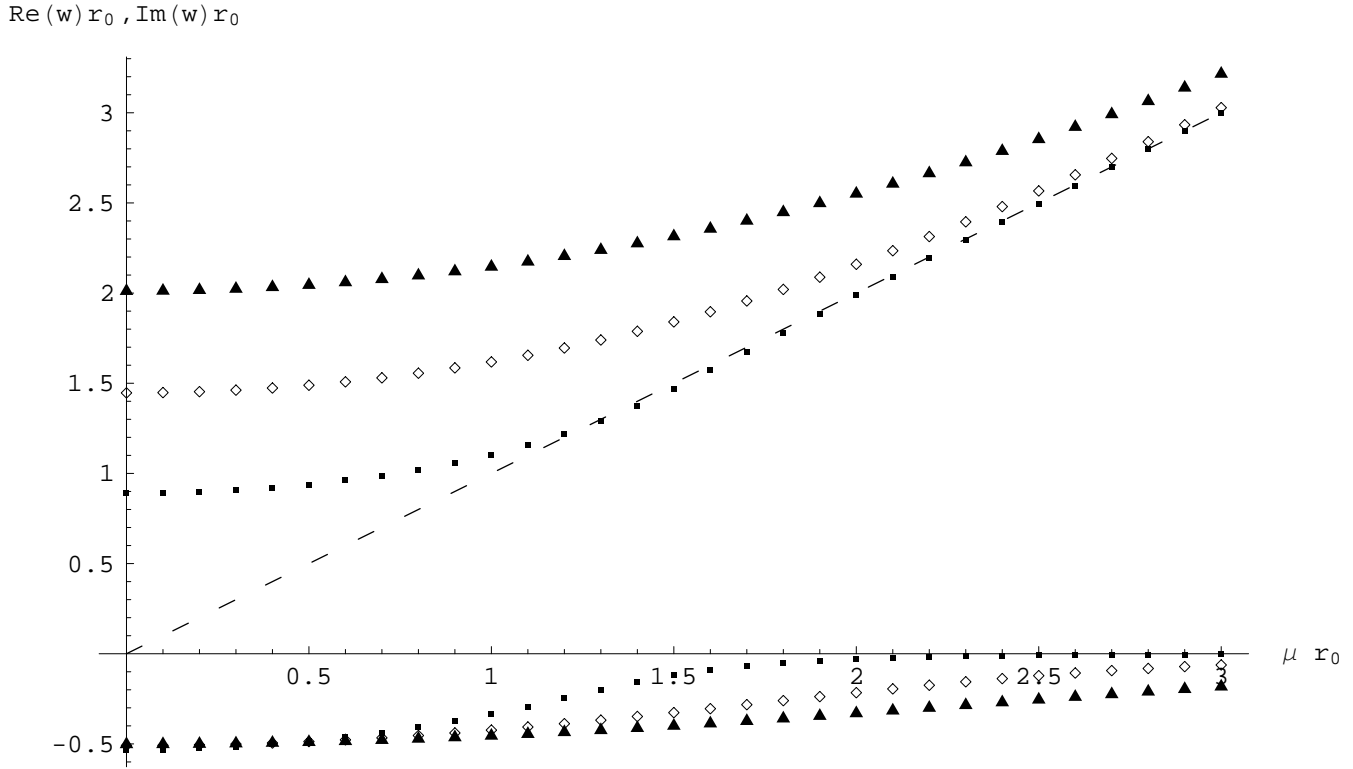}
\caption{Fundamental quasi-normal frequencies for $D=6$ as function of $\mu r_0$ for $l=0,1,2$ (represented as dots, rhombuses and triangles respectively).
The dashed line corresponds to $Re(\omega)=\mu$.}
\label{fig.D=6}
\end{center}
\end{figure*}

\begin{figure*}
\begin{center}
\includegraphics[width=\textwidth, clip]{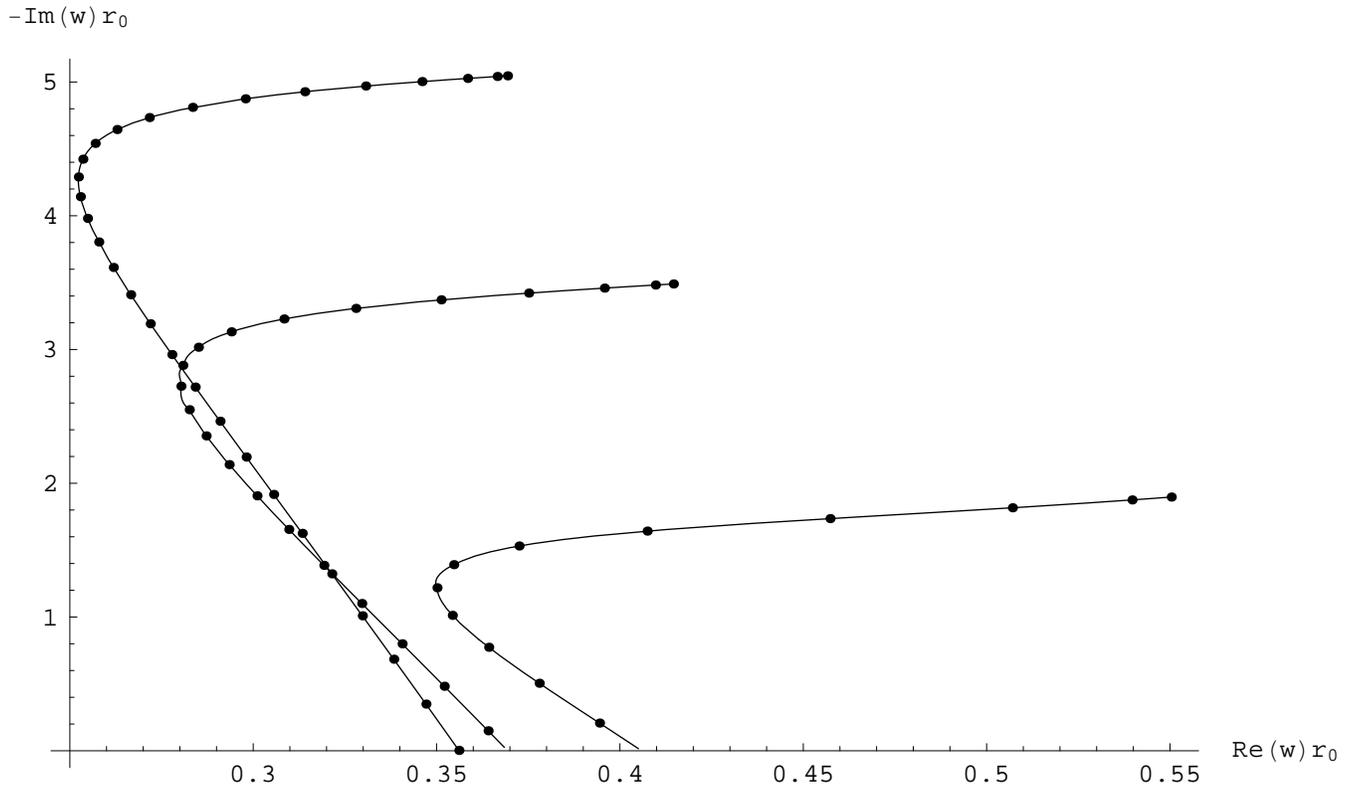}
\caption{Three higher quasi-normal modes ($l=0$) for $D=6$ and different $\mu$.
The frequency for  $\mu=0$ has the largest imaginary part. The points were plotted with the step of $\Delta\mu=1/(2r_0)$.
Solid lines mark the same overtone number.}
\label{fig.D=6.high}
\end{center}
\end{figure*}

\begin{figure*}
\begin{center}
\includegraphics[width=\textwidth, clip]{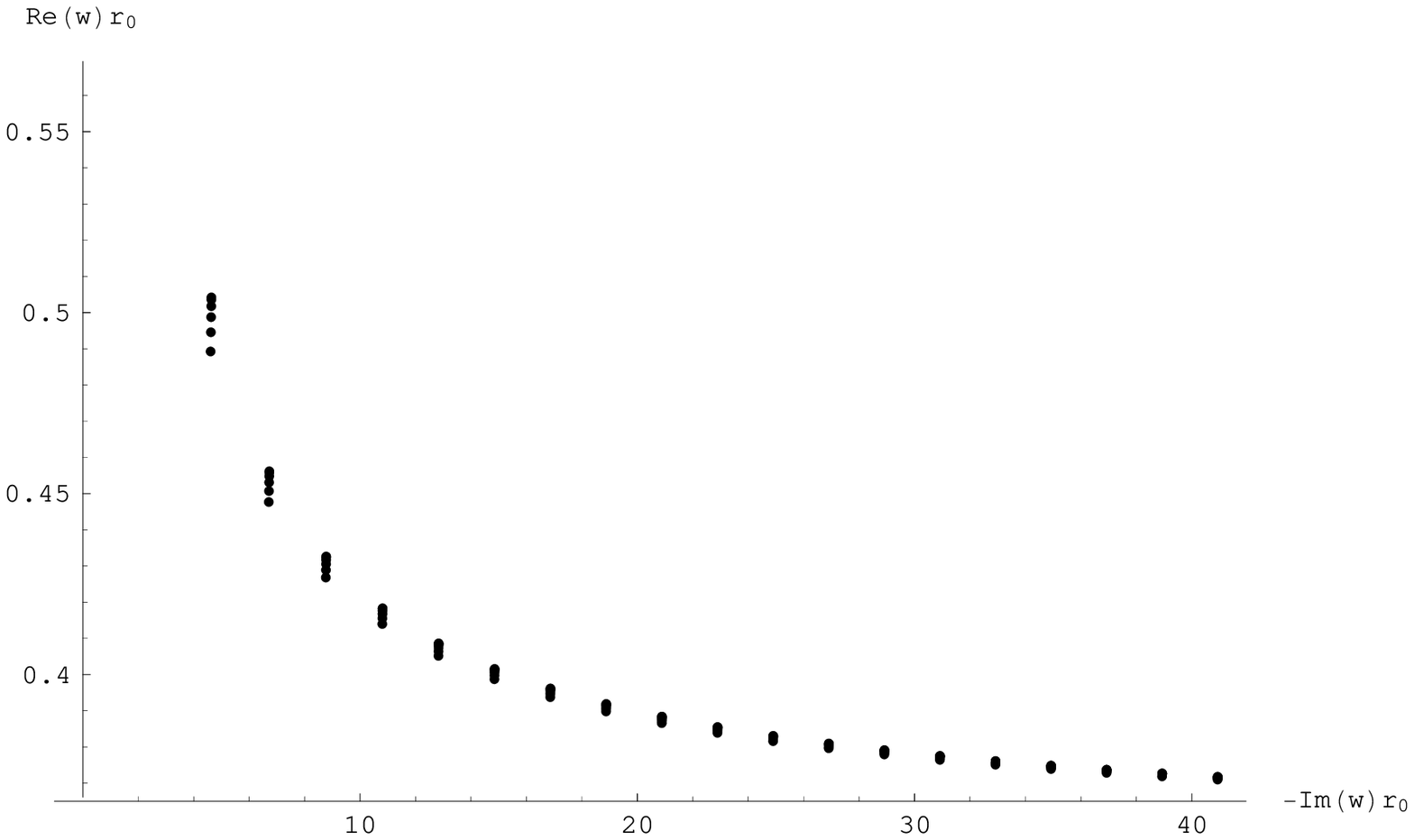}
\caption{First 20 quasi-normal modes for $D=7$, $l=0$, $\mu r_0=0,0.2,0.4,0.6,0.8,1.0$.}
\label{fig.highqnms}
\end{center}
\end{figure*}

%\end{widetext}

\section{Numerical results}\label{sec.numerical}

Let us start from an important notion.
Since the black hole mass in $D$-dimensional space-time cannot be measured in units of length, as in four-dimensional space-time,
we measure all the values in units of the black hole horizon radius that is an observable quantity.
In order to do this we multiply (\ref{radial}) by $r_0^2$ and consider dimensionless quantities $\mu r_0$ and $\omega r_0$.

Making use of the technique described in the previous section, we have found dependence of the quasi-normal modes on the field mass.
From the figure~\ref{fig.l=0} we see that for four- and five-dimensional cases increasing of $\mu r_0$ gives rise
to decreasing of the imaginary part of the fundamental quasi-normal frequency until reaching zero, while the real part changes insignificantly.
As discussed in \cite{qrc}, when some threshold value of $\mu r_0$ is exceeded, the particular quasi-normal mode disappears.
The higher threshold value corresponds to the higher overtone number.

Since we reached purely real frequency numerically with some finite accuracy, we cannot state that purely real frequencies exist in the spectrum.
We are on position, that at least while $\mu r_0$ tends to some threshold values, the quasi-normal ringing lifetime increases infinitely.

For $D\geq6$ we observe that the imaginary part of the fundamental mode tends to zero asymptotically while the real part approaches $\mu$.
We observed the same behaviour for different multipole numbers.
For $D=6$ it is shown on the figure~\ref{fig.D=6}.
The picture is similar for higher dimensional cases,
but since massless field quasi-normal frequency has higher imaginary part,
one needs to reach higher $\mu r_0$ to see this tendency.

Even though the behaviour of the fundamental overtone is different for $D\geq6$,
the dependence on $\mu r_0$ of higher overtones is qualitatively the same (see figure~\ref{fig.D=6.high}) as for $D=4$ and $D=5$.
Also because all the equations depend on $\mu^2$ we observe that the contribution of mass for higher overtones is of order $\mu^2/\omega$ as in four-dimensional case \cite{qrc,mkerr}.
Thus the asympotical behaviour of the high overtones does not depend on $\mu$.
From the figure~\ref{fig.highqnms} one can see that for 7-dimensional case the difference between massless and massive $\mu=r_0^{-1}$ cases is insignificant already for 20th overtone.
According to the analytical formula \cite{HDASQNM}, the spectrum is asymptotically equidistant for all $D$:
$$\Delta\omega=\frac{\imo}{2} f'(r_0)$$ and the real part approaches $$Re(\omega)=\frac{1}{4\pi}f'(r_0)\ln3=T_{Hawking}\ln3.$$

The difference in the mass dependence of the fundamental overtone for $D\geq6$ leads to another remarkable fact.
Since for high field mass the imaginary part of the fundamental overtone tends to zero only asymptotically (see figure~\ref{fig.l=0}),
and the imaginary part of higher overtones reaches zero for some finite value of $\mu r_0$ (see figure~\ref{fig.D=6}),
there are some values of $\mu r_0$ for which the imaginary parts of two overtones \textit{are the same}.
After one of these values is reached the overtones can be distinguished only by their real part
and the quasi-normal ringing has two dominant frequencies in its spectrum.

It turns out that this could happen only for relatively large field mass, thus the imaginary part of those frequencies is very small:
for example, for the field configuration of spherical symmetry ($l=0$) in $D=6$, the smallest value of $\mu r_0\approx5.823$,
for which there are dominant modes with $\omega r_0\approx 5.823$ and $\omega r_0\approx 0.406$.
Their imaginary parts are equivalently small, thus their lifetime is extremely high.
Thus one could observe the superposition of that two frequencies at late times of the quasi-normal ringing.

One should note that this exotic behaviour was found for relatively large field mass
for which the linear approximation (i.\ e.\ without taking into consideration the back reaction of the field upon the black hole)
could not be good.
%Note, that since $\mu r_0$ could also be large due to $r_0$, we might observe this phenomenon for large black holes if our universe is at least six-dimensional.

The fundamental modes of massless scalar field for different $D$ are presented in table~\ref{fundamentalmodes}.
We see that as $D$ increases, the real part of the fundamental overtone increases faster than the imaginary part.
Thus fields that live in the bulk are better oscillators than brane-localised ones \cite{KK}, considered in the context of Large Extra Dimension Scenario (see \cite{Kanti} for review).

\begin{table}
\caption{Fundamental quasi-normal frequencies of massless scalar field (measured in the units of the black hole horizon radius).}\label{fundamentalmodes}

\smallskip

\begin{tabular}{|l|c|c|c|}
\hline
$D$&$l=0$&$l=1$&$l=2$\\
\hline
$4$&$0.22091-0.20979\imo$&$0.58587-0.19532\imo$&$0.96729-0.19352\imo$\\
$5$&$0.53384-0.38338\imo$&$1.01602-0.36233\imo$&$1.51057-0.35754\imo$\\
$6$&$0.88944-0.53310\imo$&$1.44651-0.50927\imo$&$2.01153-0.50194\imo$\\
$7$&$1.27054-0.66578\imo$&$1.88140-0.64108\imo$&$2.49678-0.63188\imo$\\
$8$&$1.66879-0.78557\imo$&$2.32073-0.76102\imo$&$2.97469-0.75056\imo$\\
$9$&$2.07943-0.89520\imo$&$2.76402-0.87137\imo$&$3.44882-0.86013\imo$\\
\hline
\end{tabular}
\end{table}

We checked the correctness of our results by comparing them with those \cite{WKBcheck} obtained with the help of the WKB method \cite{will} for higher values of the multipole number.
The WKB method is known to give very accurate results in this regime \cite{WKBaccuracy}.

One should note also, that massless scalar field quasi-normal spectrum coincides with that of the metric perturbation of the tensor type for $l\geq2$,
which was recently studied \cite{Rostworowski:2006bp}. Our results were obtained independently and demonstrate complete agreement.

\section{Conclusions}\label{sec.conclusions}
In the present paper we studied quasi-normal spectrum of the massive scalar field near the higher dimensional black hole
within continued fraction method of Leaver with improvement of Nollert
that was generalised for the case of $k$-term recurrence relation for Frobenius series coefficients.

We found that the fundamental modes for $D\geq6$ have qualitatively different dependence on the field mass:
their real part approaches the field mass while the imaginary part tends to zero for high values of $\mu r_0$.
At the same time, the behaviour of higher modes is qualitatively the same as for $D=4$ and $D=5$ cases:
the imaginary part decreases quickly and the particular mode disappears after reaching some threshold value of $\mu r_0$.

These cause a very interesting result:
for some particular mass, the spectrum has two dominating oscillations with an identical very long lifetime.

As in four-dimensional case asymptotically high overtones do not depend on the field mass and satisfy the analytical formula for the massless case:
$$\omega_n=T_{Hawking}(\ln3-2\pi\imo n).$$

Also we found that despite the lifetime of the oscillations is shorter for larger $D$,
the increasing of the number of the extra dimensions causes black holes to be better oscillators.

\begin{acknowledgments}
I would like to acknowledge R. A. Konoplya for useful discussions
and reading this manuscript.

The work was supported by \emph{Funda\c{c}\~{a}o de Amparo
\`{a} Pesquisa do Estado de S\~{a}o Paulo (FAPESP)}, Brazil.

\end{acknowledgments}

%%%%%%%%%%%%%%%%%%%%%%%%%%%%%%%%%%%%%%%%%%%%%%%%%%%%%%%%%%%%%%%%%%%%%%%%%%%%%%%

\end{document}